\documentclass{aa}  
\usepackage{graphicx}
\usepackage{txfonts}
\begin{document}
   \title{Connecting RS OPh to [some] Type Ia Supernovae}

   \author{F. Patat
          \inst{1},
          N.N. Chugai\inst{2},
          Ph. Podsiadlowski\inst{3},
          E. Mason\inst{4,5},
          C. Melo\inst{4}
          \and
          L. Pasquini \inst{1}\fnmsep\thanks{Based on observations obtained
	  at ESO-La Silla.}
          }

   \offprints{F. Patat}
   \institute{European Organization for Astronomical Research in the
	Southern Hemisphere; K. Schwarzschild-Str. 2, 85748 Garching, Germany\\
              \email{fpatat@eso.org}
         \and
              Institute of Astronomy RAS, 48 Pyatnitskaya Str.,119017, 
              Moscow, Russia
         \and
           Department of Astrophysics, University of Oxford, 
           Oxford OX1 3RH, UK
         \and
           European Organization for Astronomical Research in the
	Southern Hemisphere; Alonso de Cordova 3107, Casilla 19001 
        -Santiago 19, Chile
        \and
           European Space Agency-Space Telescope Science Institute, 
           3700 San Martin Drive, Baltimore, MD 21218, USA
             }

   \date{Received November 2010}

 
  \abstract
   {}
   {Recurrent nova systems like RS Oph have been proposed as a possible 
   channel to Type Ia Supernova explosions, based on the high mass of
   the accreting white dwarf. Additional support to this hypothesis has been 
   recently provided by the detection of circumstellar material around 
   SN~2006X and SN~2007le, showing a structure compatible with that expected 
   for recurrent nova outbursts.
   In this paper we investigate the circumstellar environment of RS Oph 
   and its structure, with the aim of establishing a firmer and independent
   link between this class of objects and Type Ia SN progenitors.}
   {We study the time evolution of \ion{Ca}{ii}, \ion{Na}{i} and \ion{K}{i} 
   absorption features in RS Oph, before, during, and after the last outburst,
   using multi-epoch, high-resolution spectroscopy, and applying the same 
   method adopted for SN~2006X and SN~2007le.}
   {A number of components, blue-shifted with respect to the systemic velocity
   of RS Oph, are detected. In particular, one feature strongly weakens
   in the first two weeks after the outburst, simultaneously with
   the disappearance of very narrow P-Cyg profiles overimposed on the much
   wider nova emission lines of H, He, \ion{Fe}{ii} and other elements. 
 } 
   {We interpret these facts as the signature of density enhancements in the
   circumstellar material, suggesting that
   the recurrent eruptions might indeed create complex structures within the
   material lost by the donor star. This establishes a strong link between 
   RS Oph and the progenitor system of the Type Ia SN~2006X, 
   for which similar features have been detected.}


\authorrunning{F. Patat et al.}

   \maketitle
%

\section{Introduction}
\label{sec:intro}

RS Ophiuchi is a symbiotic recurrent nova (Kenyon \cite{kenyon}), that
has undergone a number of recorded eruptions, with an inter-outburst
period of 10-20 years. The binary system is composed by a K4-M0 red
giant (RG) star in a 456 day orbit with a white dwarf (WD) of mass
close to the Chandrasekhar limit (Dobrzycka \& Kenyon \cite{danuta};
Shore et al. \cite{shore}; Fekel et al. \cite{fekel}; Brandi et
al. \cite{brandi}). For this reason, RS Oph has been proposed as the
prototype for one of the possible channels to Type Ia Supernova
explosions (see for example Hachisu \& Kato
\cite{hachisu00,hachisu01}; Hernanz \& Jos\'e \cite{hernanz}). The WD
is supposed to accrete part of the material lost by the RG via stellar
wind, until it reaches the conditions for thermonuclear burning at its
surface. When this happens, the WD ejects 10$^{-8}$ to 10$^{-6}$
M$_\odot$ at velocities of $\sim$5,000 km s$^{-1}$ (Hachisu \& Kato
\cite{hachisu01}; Yaron et al. \cite{yaron}), producing a shock wave
that propagates into the pre-existing RG wind and gives rise to strong
X-ray emission, as observed in the 1985 eruption (Mason et
al. \cite{mason87}).  The last outburst of RS Oph took place on 2006
February 12.83 UT (Narumi et al. \cite{narumi}), and prompt radio
observations have shown that the nova ejection was bipolar (O'Brien et
al. \cite{obrien}). Besides confirming that the blast wave clearly
deviated from spherical symmetry, the X-Ray data (Bode et
al. \cite{bode06}; Sokoloski et al. \cite{sokoloski}) have indicated a
low mass for the ejected material which, in turn, favors a high mass
for the WD ($\sim$1.4 M$_\odot$), giving further support to the
proposed connection between RS Oph and Type Ia progenitors (Sokoloski
et al. \cite{sokoloski}). One important question that remains
  unanswered, is whether the hot component of RS Oph is a O-Ne-Mg WD
  rather than a C-O WD, for there are no direct observations that
  could help to distinguish between the two. In this case, the WD
  would not experience a thermonuclear explosion, leading to a SN Ia,
  but rather undergo an electron-capture core collapse, producing a
  neutron-star remnant (Nomoto \& Kondo \cite{nomoto}), when its mass
  gets close to the Chandrasekhar mass.

In addition to the clear detection of circumstellar matter (CSM) very
close to the system (O'Brien et al.  \cite{obrien}; Sokoloski et al.
\cite{sokoloski}), Iijima (\cite{iijima,iijima09}) has reported the
presence of a narrow \ion{Na}{i} D absorption component, which is
blue-shifted by about 33 km s$^{-1}$ with respect to the systemic
velocity of RS Oph. He proposed that this feature might arise in the
wind of the RG, hence arguing for its circumstellar nature. Stimulated
by this finding and by the detection of \ion{Na}{i} D time-variant
features in the Type Ia SN~2006X (Patat et al. \cite{patat}),
SN~2007le (Simon et al. \cite{simon}), 1999cl (Blondin et
al. \cite{blondin}), and SN~2006dd (Stritzinger et al. \cite{max}) we
performed a thorough analysis using pre/post-outburst multi-epoch,
high-resolution spectroscopy of RS Oph, whose results we are reporting
here.


\section{Observations and Data Reduction}
\label{sec:obs}

RS Oph was observed in the pre-outburst, outburst and post-outburst
phases using the Fiber-fed Extended Range Optical Spectrograph (FEROS;
Kaufer et al. \cite{kaufer}) mounted at the ESO-MPG 2.2m
telescope. The pre-outburst data set was presented by Zamanov et
al. (\cite{zamanov}), while the outburst data are going to be
thoroughly discussed in a forthcoming paper, to which we refer the
reader for a full report on the high-resolution spectroscopic
follow-up of RS Oph (Mason et al. in preparation). In order to cover
the quiescence phase following the outburst, an additional set of data
was obtained in February 2008, about two years after the 2006
eruption.

The instrument delivers a resolving power $\lambda/\Delta \lambda
\sim$ 48,000, covering the spectral range 3700--9200 \AA. The data
were processed by the FEROS pipeline, which includes the correction
for Earth's motion. The first set of data was obtained on four epochs
between 672 and 530 days before the February 2006 outburst, whilst the
second covers 5 epochs between 1.5 and 46.6 days after the
eruption. Finally, RS Oph was observed on three consecutive nights
about 740 days after the eruption, when the system was well back into
the quiescence phase, and the mass accretion had resumed (Worters et
al. \cite{worters}). The spectra obtained on these three nights were
stacked in order to increase the signal-to-noise ratio. The resulting
spectrum is presented and discussed here for the first time.

A journal of the observations is given in Table~\ref{tab:obs}. The
presence and position of telluric lines, produced by H$_2$O and O$_2$
absorption bands, was checked using a synthetic spectrum computed
by the Reference Forward
Model\footnote{http://www.atm.ox.ac.uk/RFM/}. Gas column densities
were estimated using VPGUESS\footnote{VPGUESS was developed by
  J. Liske and can be downloaded at
  http://www.eso.org/$\sim$jliske/vpguess/index.html}, and refined
with VPFIT\footnote{VPFIT was developed by R.F. Carswell and can
  be downloaded at http://www.ast.cam.ac.uk/$\sim$rfc/vpfit.html}.

\begin{table}
\caption{\label{tab:obs}Log of RS Oph FEROS observations.}
\centerline{
\tabcolsep 0.6mm
\begin{tabular}{cccccccc}
\hline
Date        & MJD & Phase & $r_v$ & Epoch  & Airm. & Exp. & Bar.Corr.\\
            &    &    & (km s$^{-1}$) & (days) & & (s) & (km s$^{-1}$) \\     
\hline
2004-04-11 & 53106.39 & 0.48 &$-$16.6 & $-$672.4  & 1.08 & 1200 & +26.1\\   
2004-06-05 & 53162.23 & 0.60 &$-$13.5 & $-$616.6  & 1.09 & 1200 & +6.3 \\   
2004-06-06 & 53163.29 & 0.60 &$-$13.3 & $-$615.5  & 1.15 & 1200 & +5.7 \\     
2004-08-31 & 53248.05 & 0.79 &+4.1   & $-$530.8  & 1.13 & 1200 & $-$26.9\\ 
\hline
2006-02-14 & 53780.37 &	0.96 &+16.1 &+1.54     & 2.09 &  120 & +24.7\\   
2006-02-25 & 53791.37 &	0.98 &+16.6 &+12.54    & 1.57 &   30 & +27.1\\   
2006-03-07 & 53801.40 &	0.00 &+16.7 &+22.57    & 1.22 &   90 & +28.4\\   
2006-03-18 & 53812.42 &	0.03 &+16.5 &+35.59    & 1.10 &  180 & +28.7\\   
2006-03-31 & 53825.41 & 0.06 &+15.7 &+46.58    & 1.08 &  300 & +27.9\\  
\hline
2008-02-24 & 54520.35 & 0.58 &$-$14.5&+741.52  & 1.72 & 4840 & +26.9\\
2008-02-25 & 54521.35 & 0.58 &$-$14.5&+742.52  & 1.75 & 4200 & +27.0\\
2008-02-26 & 54522.36 & 0.58 &$-$14.5&+743.53    & 1.78 & 4650 & +27.2\\
\hline
\end{tabular}
} Note: Epochs are computed from the 2006 outburst (February 12.83 UT,
Narumi et al. \cite{narumi}). Phase and radial velocities for the M
giant ($r_v$) are computed using the ephemeris by Fekel et
al. (\cite{fekel}).

\end{table}

\section{\label{sec:islines}Inter- and/or circumstellar lines towards RS Oph}

\begin{figure}
\centerline{
\includegraphics[width=9cm]{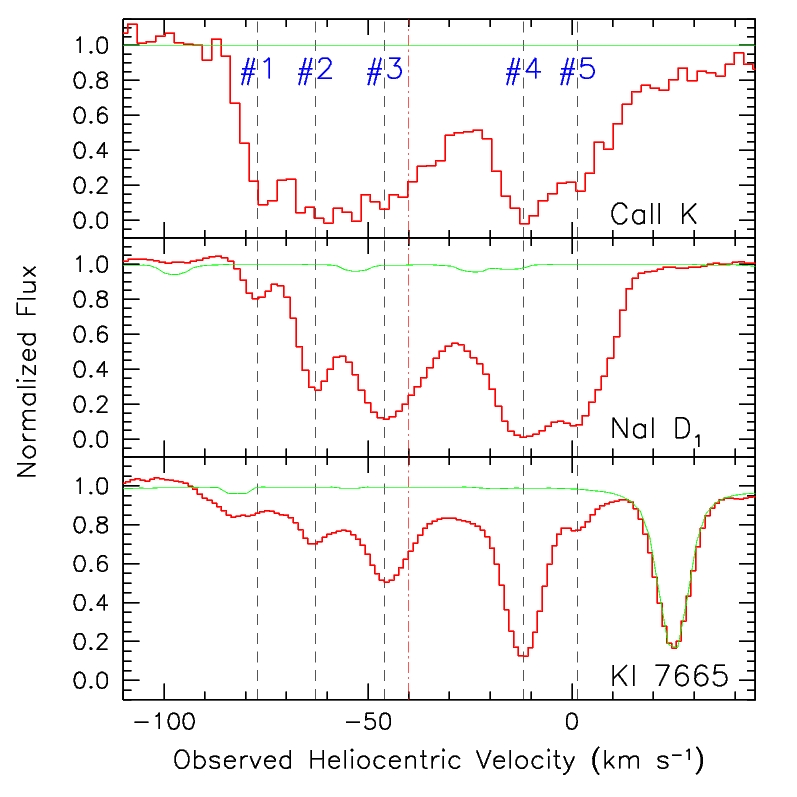}
}
\caption{\label{fig:islines}Interstellar lines along the line-of-sight
  to RS Oph: \ion{Ca}{ii} K (upper), \ion{Na}{i} D$_1$ (middle), and
  \ion{K}{i} $\lambda$7665 (lower) on day $-$672. The vertical dashed
  lines mark the main components (see text). The dashed-dotted line
  indicates the systemic velocity ($-$40.2 km s$^{-1}$; Fekel et
  al. \cite{fekel}). The green curve is a synthetic telluric
  absorption spectrum matched to the observed spectrum.}
\end{figure}

Pronounced \ion{Ca}{ii} H\&K, \ion{Na}{i} D, \ion{K}{i} $\lambda$7665,
and \ion{K}{i} $\lambda$7699 are detected in the high-resolution
spectra of RS Oph, as it is shown in Fig.~\ref{fig:islines}. While the
\ion{K}{i} $\lambda$7699 profiles are rather severely affected by
telluric absorptions, these are practically negligible for the other
lines. Besides two intense features at heliocentric velocity $-$12 km
s$^{-1}$ (\#4) and +2 km s$^{-1}$ (\#5), arising in the local
spiral arm (Hjellming et al.  \cite{hjellming}), the data reveal at
least three distinct components\footnote{In the spectra presented by
  Iijima (\cite{iijima,iijima09}), obtained on February 18.2 (+5.4)
  and 19.2 (+6.4) with a resolution $\lambda/\Delta
  \lambda\sim$10,000, these components are severely blended.} at $-$77
(\#1), $-$63 (\#2) and $-$46 km s$^{-1}$ (\#3), respectively. The same
components, though with different intensities, are visible in all
\ion{Ca}{ii}, \ion{Na}{i}, and \ion{K}{i} features
(Fig.~\ref{fig:islines}). Besides those already mentioned, other
shallower features are most likely present between $-$40 and $-$20 km
s$^{-1}$, but they contribute only marginally to the overall profile.

As the systemic velocity of RS Oph is $-$40.2 km s$^{-1}$ (Fekel et
al. \cite{fekel}; see also Brandi et al. \cite{brandi}), components
\#1 to \#3 must arise in material coming towards the observer, at
velocities where no interstellar component is expected (Kerr \&
Westerhout \cite{kerr}; see also Cassatella et al. \cite{cassatella};
Hjellming et al. \cite{hjellming}; Snijders \cite{snijders}).

This is a most intriguing fact, especially if it can be proven that
these absorptions arise in the circumstellar environment. In this
context, the first question to be answered is about the
possible/probable presence of a photospheric absorption component,
which may further complicate the picture.

\begin{figure}
\centerline{
\includegraphics[width=9cm]{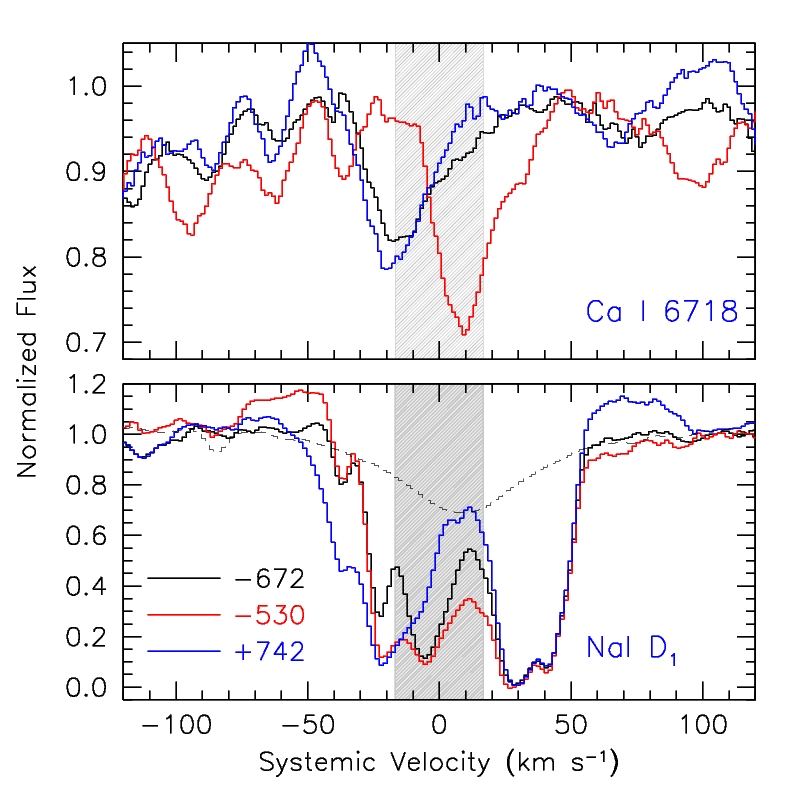}
}
\caption{\label{fig:ca}Time evolution for \ion{Na}{i} D$_1$ (lower
  panel) and \ion{Ca}{i} $\lambda$6717.7 (upper panel) on three
  quiescence epochs ($-$672, $-$530 and +742).  The vertical grey
  strip indicates the radial velocity range of the M giant during the
  orbital period ($\pm$16.2 km s$^{-1}$; Fekel et
  al. \cite{fekel}). The dashed curve in the lower panel is the line
  profile of the K giant star HD32887 (Da Silva et al. \cite{dasilva}),
  veiled to 30\% of its original depth and placed at the velocity
  deduced from \ion{Ca}{i} $\lambda$6718 for day $-$530.}
\end{figure}

\section{\label{sec:giant}The spectrum of the giant during quiescence}

The spectral type of the donor star in the RS Oph system, based on the
study of molecular bands in the near-IR, is uncertain and ranges from
K4~III to M0~III (M\"urset \& Schmid \cite{muerset}). The effective
temperature is estimated to be in the range 4100--4400 K, and the
metallicity is [M/H]=0.17$\pm$0.1. Additionally, the star is reported
to show an anomalously high Li abundance (Wallerstein et
al. \cite{wallerstein}; see also the discussion in Hernanz \& Jos\'e
\cite{hernanz} and Brandi et al. \cite{brandi}).

Because of the presence of a hot component, the spectrum of the giant
is heavily veiled (Bruch \cite{bruch}), so that below 4000 \AA\/ this
is dominated by the hot component, while at 5500 \AA\/ the
contribution of the cool component is between 30 and 50\% (Dobrzycka
et al. \cite{danuta96}). For these reasons it is natural to expect the
strong and wide photospheric \ion{Na}{i} D lines typical of a cool giant to
be visible in the composite spectrum\footnote{Marked \ion{Na}{i} D
  lines, supposedly arising from the cool component, were reported by
  Bruch (\cite{bruch}), with a global equivalent width of about 3
  \AA. Nevertheless, given the low resolution of Bruch's data, most of
  the absorption he measured is most probably caused by the interstellar
  component.}.

\begin{figure}
\centerline{
\includegraphics[width=9cm]{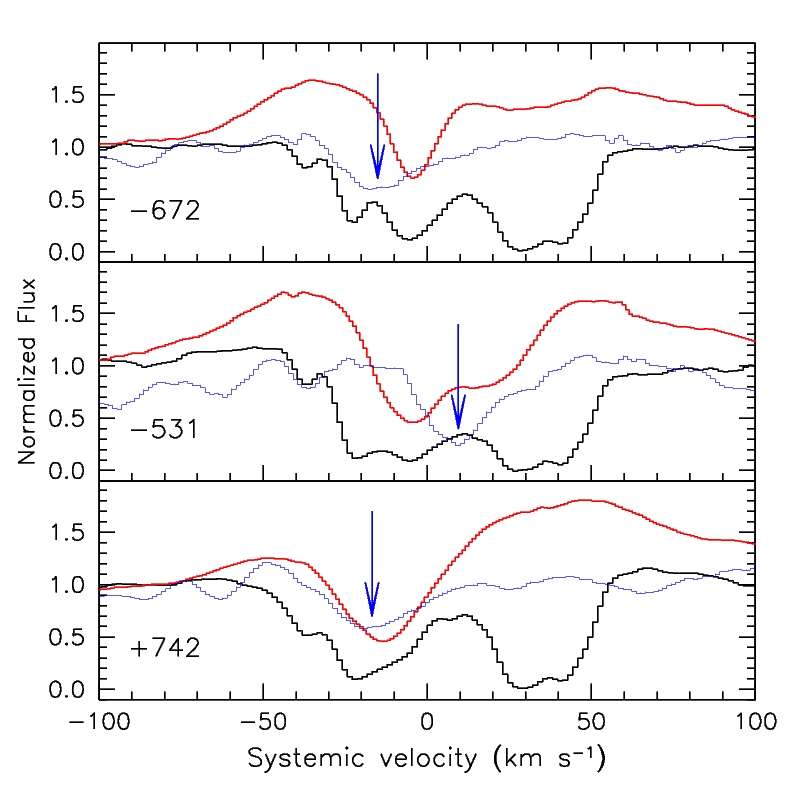}
}
\caption{\label{fig:ircana}Comparison between the evolution of \ion{Ca}{ii}
  8662.14 (red) and \ion{Na}{i} D$_1$ (black) during three quiescence
  epochs. The blue curve traces the \ion{Ca}{i} $\lambda$6718 line,
  while the blue arrow marks the estimated M giant radial velocity.}
\end{figure}

Therefore, a possible explanation for the changes in the \ion{Na}{i} D
absorption features during the quiescent phase (see Fig.~\ref{fig:ca},
lower panel) is the variation of the lines in the M giant
spectrum. These might be explained in terms of wavelength shifts due
to the orbital motion, but also as intrinsic line variability.  In
fact, the hot components of symbiotic systems are known to heat the
facing hemispheres of the red companion, hence producing temperature
changes in their atmospheres large enough to change the lines shapes
(see Kenyon et al. \cite{kenyon93} and references therein). Indeed, in
the case of RS Oph both mechanisms might be at work.  This is shown in
the upper panel of Fig.~\ref{fig:ca}, where we plotted the region of
the \ion{Ca}{i} $\lambda$6718 line for the first ($-$672 days) and the
last ($-$530 days) pre-outburst epochs in our data set. Besides the
velocity shift observed between the two epochs ($\sim$25 km s$^{-1}$),
the Ca~I line shows a significant depth increase, a behavior that is
common also to other stellar lines (like \ion{Li}{i} $\lambda$6707,
very neatly detected in our spectra).  This might explain, at least
partially, the evolution of the \ion{Na}{i} D absorption profiles
observed during the quiescence phase (Fig.~\ref{fig:ca}, lower
panel). 

An important fact to be noticed is that \ion{Na}{i} D and \ion{Ca}{ii}
H\&K regions are most probably affected by the presence of components
arising in the circumbinary environment, which are characterized by
the complex profiles seen in other prominent lines like H and He (see
for instance Zamanov et al. \cite{zamanov}; Brandi et
al. \cite{brandi}). Because of the presence of interstellar and
possible circumstellar components, it is very difficult to distinguish
these components in \ion{Na}{i} D, \ion{Ca}{ii} H\&K, and \ion{K}{i}
lines. Nevertheless, this becomes feasible for other strong lines,
where the circum/interstellar components are absent, like the near-IR
\ion{Ca}{ii} triplet. An example is presented in
Fig.~\ref{fig:ircana}, where the evolution of \ion{Ca}{ii}
$\lambda$8662.14 is compared to the one of \ion{Na}{i} D$_1$. The
changes seen in the NIR-\ion{Ca}{ii} lines are followed by those in the
\ion{Na}{i} profile and can be explained, at least partially, by the
radial velocity shift of the M giant, traced by the \ion{Ca}{i}
$\lambda$6718 line.

Therefore, during quiescence there certainly is a stellar
component, possibly characterized by a complex structure, which
affects the observed profile of the narrow absorption lines arising in
the inter/circumstellar environment.

\section{Evolution of absorption features during outburst}
\label{sec:evol}

\begin{figure}
\centerline{
\includegraphics[width=9cm]{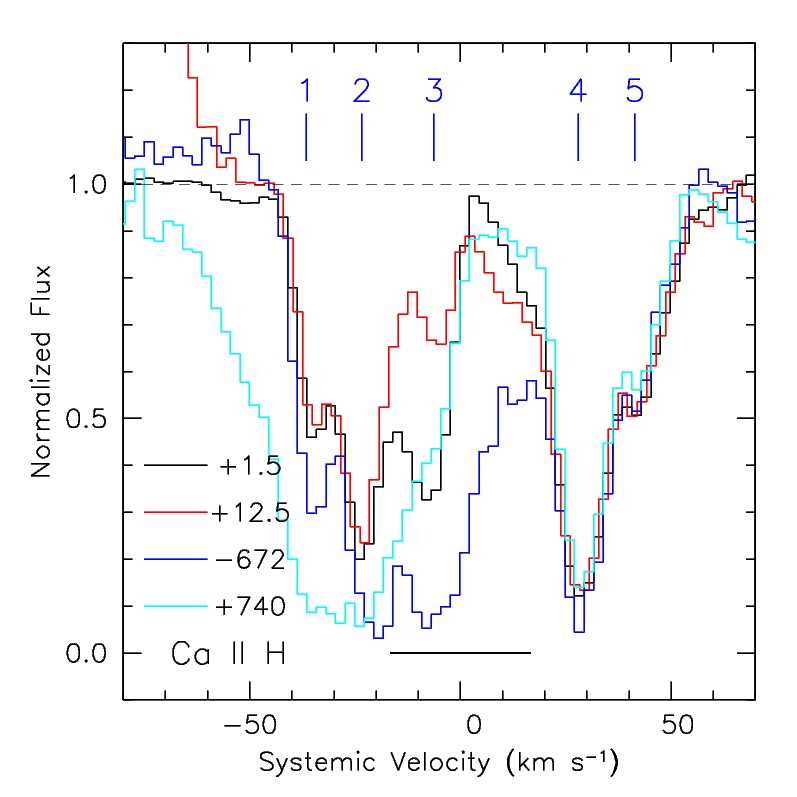}
}
\caption{\label{fig:caevol} Evolution of the \ion{Ca}{ii} H profile on
  days $-$672, +1.5, +12.5 and +740. The horizontal bar indicates the
  radial velocity range of the M giant during a full orbit.}
\end{figure}

\begin{figure}
\centerline{
\includegraphics[width=9cm]{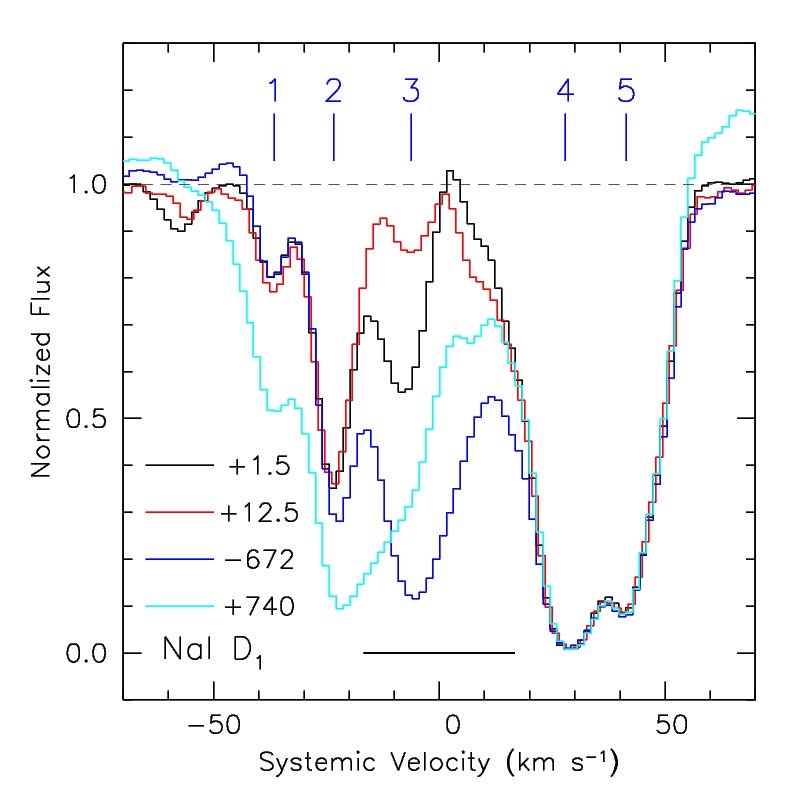}
}
\caption{\label{fig:naevol} Same as Fig.~\ref{fig:caevol} for
  \ion{Na}{i} D$_1$.}
\end{figure}

\begin{figure}
\centerline{
\includegraphics[width=9cm]{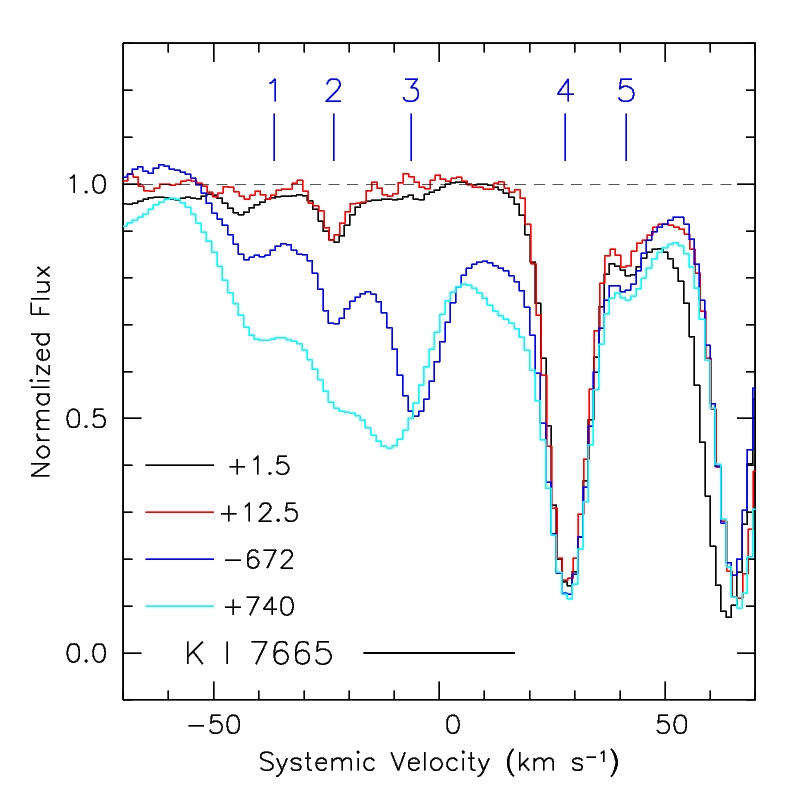}
}
\caption{\label{fig:kievol} Same as Fig.~\ref{fig:caevol} for
  \ion{K}{i} $\lambda$7665. The absorption at $\sim$+65 km s$^{-1}$ is
  a telluric feature.}
\end{figure}

The time evolution during one pre-outburst ($-$672 days), two outburst
(+1.5, +12.5 days), and one post-outburst (+742 days) epochs is
presented in Figs.~\ref{fig:caevol}, \ref{fig:naevol},
\ref{fig:kievol} for \ion{Ca}{ii} H, \ion{Na}{i} D$_1$, and
\ion{K}{i}, respectively. The most striking fact shown by these plots
is the marked evolution of component \#3.

For the specific case of \ion{Na}{i}, in the pre-outburst epoch this
component is almost completely saturated, while on day +1.5 its depth
has decreased by about a factor 2. The line appears to have weakened
further on day +12.5, after which it does not show any significant
evolution up to the last epoch covered by our data (+46.6). The column
density retrieved from component \#3 changes from $\log N\sim$12.9 on
day $-$672 to $\sim$12.1 on day +1.5, to finally reach $\sim$11.3 on
day +22.6, implying a variation of over a factor 30. The line width
($b$=FWHM/1.665) derived for this component ranges between 3 and 4 km
s$^{-1}$ during the post-outburst epochs, while for the pre-outburst
epoch the best fit gives $b\simeq$8.5 km s$^{-1}$. This is a further
indication that an additional component is present prior to outburst,
most likely due to the M giant intrinsic absorption (see
Sect.~\ref{sec:giant}).

\begin{figure}
\centerline{
\includegraphics[width=9cm]{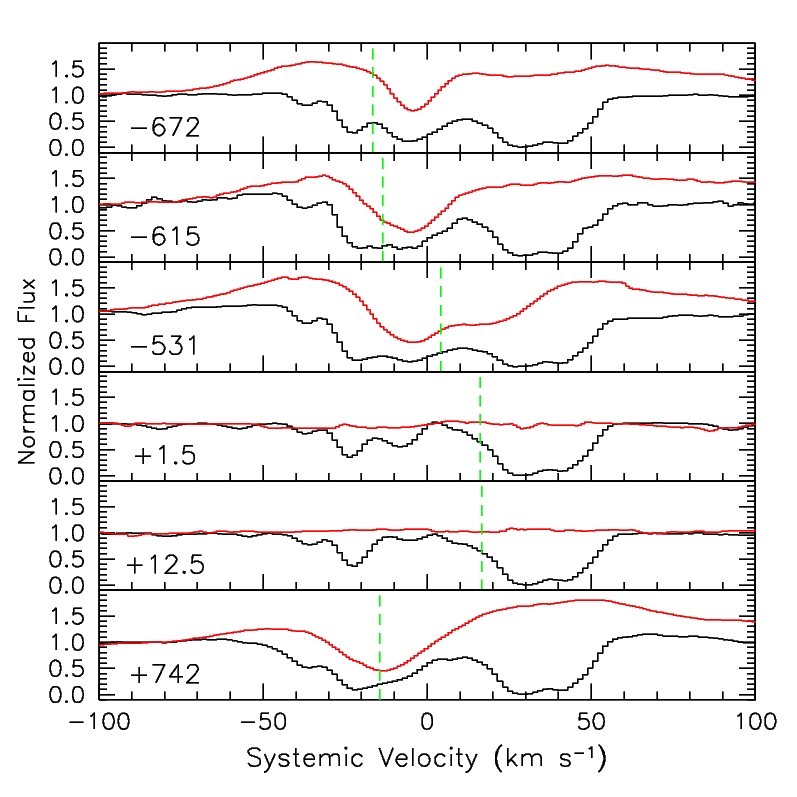}
}
\caption{\label{fig:ircana3}Comparison between the evolution of
  \ion{Ca}{ii} 8662.14 \AA\/ (red) and \ion{Na}{i} D$_1$ (black) on
  days $-$672, $-$615, $-$531, +1.5, +12.5, and +742. The dashed
  vertical lines mark the expected radial velocity of the M giant
  (Fekel et al. \cite{fekel}).}
\end{figure}

In general, the behavior seen in \ion{Ca}{ii} H\&K is very similar to
that of \ion{Na}{i} D, with component \#3 clearly decreasing in
intensity, both between the pre-outburst ($-$672) and the first
outburst (+1.5) epochs, and between the first two outburst epochs
(+1.5, +12.5). Importantly, the pronounced evolution displayed during
the first two weeks after the outburst clearly demonstrates that this
cannot be attributed solely to the disappearance of the photospheric
component. In fact, during these phases the overall luminosity of the
system is $\sim$100 times larger than that of the M giant, making its
relative contribution to the total spectrum negligible.  This is
illustrated in Fig.~\ref{fig:ircana3}, where we plotted the comparison
between the evolution of the near-IR \ion{Ca}{ii} $\lambda$8662 and
\ion{Na}{i} D$_1$ across several of the epochs covered by our
observations. While during quiescence the complex profile of
\ion{Ca}{ii} is clearly detected, this is completely missing during
the outburst. On the contrary, barring the variation seen in component
\#3, the blue-shifted \ion{Na}{i} components are always visible, ot
all epochs. The same is true for the \ion{Ca}{ii} H\&K features.

Therefore, absorption components \#1, \#2 and \#3 certainly arise
outside of the stellar atmosphere, and outside the nova ejecta. In
addition, the variation seen in component \#3 tells us that at least
that feature cannot be interstellar. In this respect, an even more
conclusive argument about the nature of all blue-shifted features
comes from a closer inspection of Fig.~\ref{fig:kievol}. At variance
with what is seen for \ion{Na}{i} D and \ion {Ca}{ii} H\&K, the
\ion{K}{i} absorptions disappear almost completely during the outburst
(component \#2 is still visible on days +1.5 and +12.5. See also
Sect.~\ref{sec:disc}). This definitely rules out an interstellar
origin for these lines.

A hint to the physical mechanism responsible for the observed line
variations comes from the following consideration. The variations
appear to be stronger for species with lower ionization potentials:
the largest changes are seen for \ion{K}{i} (4.3 eV), while the
weakest ones are observed for \ion{Ca}{ii} (11.9 eV), with \ion{Na}{i}
(5.1 eV) showing an intermediate behavior. This suggests that the
weakening of absorption features is related to a change in the
ionization balance, induced by the radiation field produced by the
nova and/or by the interaction between the nova ejecta and
pre-existing, circumstellar material.

At variance with the SN case, where the system is completely disrupted
by the explosion, the recurrent nova case offers the possibility of
studying the CS environment also after the outburst. With the aid of
the +742 epoch, obtained when the system is supposed to be back to the
quiescence phase (Worters et al. \cite{worters}), one can directly
verify whether the outburst has modified the CSM in a sensible way. In
this respect, the evolution shown in Figs.~\ref{fig:caevol},
\ref{fig:naevol}, and \ref{fig:kievol} is revealing. Although the
post-outburst epoch was obtained at a similar orbital phase of the
pre-outburst epoch $-$672 (see Table~\ref{tab:obs}), the profile is
markedly different, with an enhanced absorption in the blue. This is
common to \ion{Ca}{ii}, \ion{Na}{i}, and \ion{K}{i}, pointing to a
global modification of the circumstellar environment produced by the
nova ejecta.

\begin{figure}
\centerline{
\includegraphics[width=9cm]{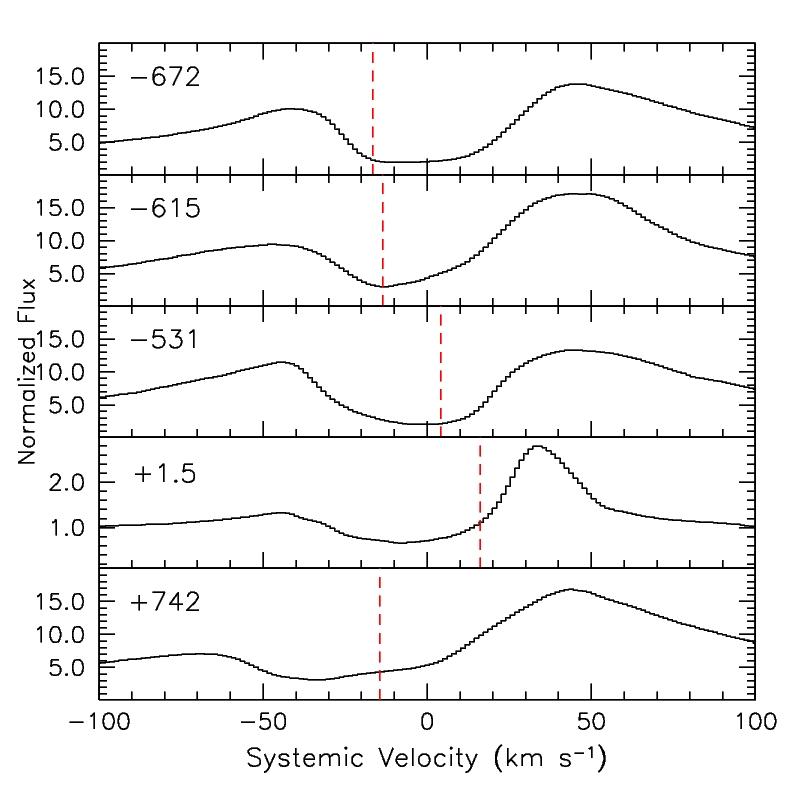}
}
\caption{\label{fig:hevol}Evolution of the H$\alpha$ profile on days $-$672,
$-$615, $-$531, +1.5, and +742 (from top to bottom). The dashed
  vertical lines mark the expected radial velocity of the M giant
  (Fekel et al. \cite{fekel}).}
\end{figure}

This conclusion is strengthened by the analysis of the H$\alpha$
profile presented in Fig.~\ref{fig:hevol}. The post-outburst profile
departs from that seen during the 3 pre-outburst epochs, in that the
absorption extends far more into the blue, reaching a systemic
velocity of about $-$60 km s$^{-1}$. Given the evolution shown in the
pre-outburst phases, it is evident that the variation seen on day +742
cannot be explained in terms of the pure orbital motion of the M
giant. The question as to whether this is a transient phenomenon or a
more stable perturbation of the circumstellar configuration will have
to await for follow-up high resolution spectroscopy.

\begin{figure}
\centerline{
\includegraphics[width=9cm]{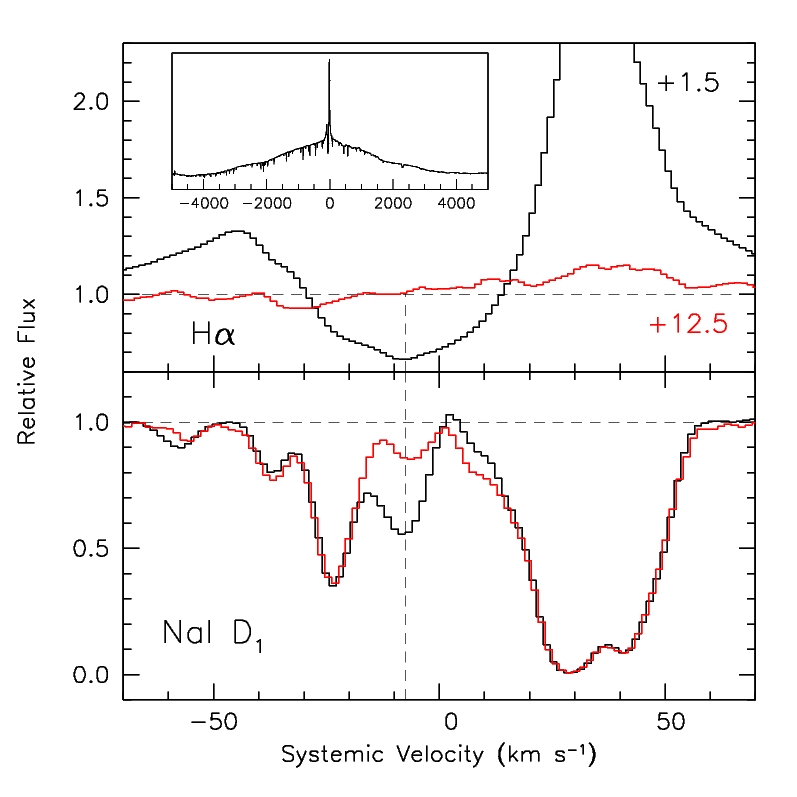}
}
\caption{\label{fig:halpha}Comparison between the evolution of H$\alpha$
(upper panel) and \ion{Na}{i} D$_1$ (lower panel) on the first two,
post-outburst epochs. The small insert in the upper panel shows the
overall profile of the H$\alpha$ line. The vertical dashed line marks
the position of the time variable \ion{Na}{i} D component \#3.}
\end{figure}

A very remarkable fact observed in conjunction with the weakening of
component \#3 is the disappearance of a narrow P-Cygni profile which
is well visible in H, He, \ion{O}{i} and basically all other broad
emission lines on day +1.5.  For example, on this epoch, H$\alpha$
shows a rather complex, multi-component profile, with a FWHM of about
3400 km s$^{-1}$, indicating exceptionally high expansion velocity for
the nova ejecta. On top of this emission, a narrow P-Cyg profile is
present, with the minimum of the absorption trough placed at a
velocity of about $-$10 km s$^{-1}$ (Fig.~\ref{fig:halpha}, upper
panel). This is notably similar to that of the time variable
\ion{Ca}{ii}/\ion{Na}{i}/\ion{K}{i} feature \#3 (see also
Fig.~\ref{fig:halpha}, lower panel). On the next epoch (day +12.5),
this narrow feature has completely disappeared, leaving the pure
underlying H$\alpha$ broad emission profile with FWHM$\sim$1250 km
s$^{-1}$.

The disappearance of the narrow P-Cyg absorptions soon after the
outburst indicates that the gas in which they originate must be close
to the eruption site, in order to be influenced by the radiation field
and/or by the direct interaction with the nova ejecta, that is known
to start immediately after the outburst (O'Brien et al. \cite{obrien};
Sokoloski et al. \cite{sokoloski}).

The most natural conclusion one can draw from all these facts is that
the narrow, blue-shifted absorptions observed in RS Oph arise in
circumstellar material.

\section{Discussion}
\label{sec:disc}

The time variations of the \ion{Na}{i}, \ion{Ca}{ii} and \ion{K}{i}
lines observed in RS Oph are rather dramatic, and cannot be compared
to the known, long timescale fluctuations observed along the line of
sight to galactic stars (Crawford \cite{crawford}). They are much more
resemblant to those observed in the Type Ia SN~2006X (Patat et
al. \cite{patat}) and 2007le (Simon et al. \cite{simon}). These were
interpreted as arising in the circumstellar material, whose physical
conditions where probably modified by the explosion, both by the
radiation field and by the direct interaction with the SN ejecta (see
the discussion in Chugai \cite{chugai08} for an alternative
scenario. See also Patat et al. \cite{patat10}). To the best of our
knowledge, short timescale variations of narrow absorptions are known
to this level of detail only for another object, i.e. the
core-collapse, type IIn SN~1998S (Bowen et al. \cite{bowen}). High
resolution spectroscopy of this object has revealed a number of
\ion{Na}{i} D components within the host galaxy, the bluest of which
deepened significantly during the 19 days spanned by the two epochs
available, implying a \ion{Na}{i} column density increase of
$\gtrsim$1 dex. SN~1998S has shown definite signs of ejecta-wind
interaction in the form of narrow H and He P-Cyg profile lines and,
remarkably, these features were detected at the same velocity of the
\ion{Na}{i} D variable component, and showed the same deepening with
time. Bowen et al. (\cite{bowen}) concluded that the narrow and
variable absorption is a {\it signature of the outflows from the
  super-giant progenitor of SN~1998S}, arising in a dense shell,
expanding at about 50 km s$^{-1}$ (see also Fassia et
al. \cite{fassia}).

Given all the evidences collected in this work (see
Sect.~\ref{sec:evol}), the most plausible possibility is that the time
variant \ion{Na}{i}, \ion{Ca}{ii}, and \ion{K}{i} features are
produced within the circumstellar environment of RS Oph. Since clear
signs of interaction between the CSM and the nova ejecta were detected
immediately after the eruption (O'Brien et al. \cite{obrien};
Sokoloski et al. \cite{sokoloski}), it is reasonable to argue that the
weakening of component \#3 is due to the interaction with the fast
moving nova ejecta, which sweep away the pre-existing material. Under
this hypothesis and given the fact that the interaction must have
already started by the time of our second outburst epoch, an upper
limit to the distance of this gas can be obtained from the ejecta
velocity ($\sim$4,000 km s$^{-1}$) and the time of interaction
($\leq$12.5 days), and it turns out to be $\leq$4$\times$10$^{14}$
cm.  

The velocity measured for the time-variable component \#3 (6.3 km
s$^{-1}$) is significantly smaller than that expected for a RG
wind. Even taking into account the combination of wind velocity,
rotational velocity, orbital velocity and hydrodynamical effects, the
radial speed of the outflowing material is expected to be between 20
and 50 km s$^{-1}$, depending on the viewing angle (Walder, Folini \&
Shore \cite{walder}). Therefore, it is plausible that component \#3
arises within the circumbinary environment, at variance with
components \#1 and \#2, which most likely arise much farther out (see
the discussion below).

The problem with this scenario is the absence of a P-Cyg profile in
the Na, Ca and K lines, contrary to what is seen in H, \ion{He}{i},
\ion{O}{i}, \ion{Fe}{ii} and \ion{Si}{ii} lines.  The most viable
explanation is that \ion{Na}{i}, \ion{Ca}{ii}, and \ion{K}{i} lines
arise in an asymmetric density enhancement lying along the line of
sight, while H and other lines are generated both in an asymmetric and
in a more symmetric material, possibly identified as a circumbinary
equatorial wind. The reason why we do not see \ion{Na}{i},
\ion{Ca}{ii}, and \ion{K}{i} lines arising from the symmetric
component is plausibly a strong ionization due to their low ionization
potential. In fact, all the ions for which P-Cyg profiles are observed
have higher ionization potentials (H: 13.6 eV; \ion{He}{i}: 24.6 eV;
\ion{O}{i}: 13.6 eV; \ion{Si}{ii}: 16.3 eV; \ion{Fe}{ii}: 16.2 eV)
than \ion{Na}{i} (5.1 eV), \ion{Ca}{ii} (11.9 eV) and \ion{K}{i} (4.3
eV) and higher (by at least 1 dex) abundance than Na, Ca and
K. Therefore the latter three species, being strongly ionized in the
wind, are practically unobservable. Only in the dense asymmetric
component the optical depth in these absorptions is sufficient to
produce detectable features. In this respect we note that the
emergence of asymmetric, dense structures lacking point-symmetry, is a
natural consequence of the orbital motion of a mass-loosing red giant
(Mastrodemos \& Morris \cite{mastrodemos}). In general, when dealing
with a system like RS Oph, one should bear in mind that the dynamic
conditions of the circumbinary gas are most likely non-stationary,
because the flow is strongly affected by the orbital motion. This may
result in the formation of transient absorbing structures in the line
of sight at similar orbital phases, which make very difficult a proper
modeling of the geometry and gas flow of the CSM. Hydrodynamical
3D-simulations by Walder et al. (\cite{walder}) show that during
quiescence the circumbinary matter of RS Oph can be described as an
oblate spheroid on the scale of $\sim10^{15}$ cm, while on smaller
scales ($\sim10^{14}$ cm) the geometry is rather complicated, and
essentially lacks point-symmetry.

As we mentioned above, the early disappearance of component \#3 is
strongly suggestive of a circumbinary nature. In this respect, it is
interesting to note that the so called cF-absorption system was
destroyed during the outburst (Brandi et al. \cite{brandi}). It then
reappeared (accompanied by flickering) about 8 months after,
indicating the resumption of accretion (Worters et
al. \cite{worters}). The cF-absorptions are believed to be related to
the hot component of the binary system (Mikolajewska \& Kenyon
\cite{mikolajewska}). In addition, their measured velocities
are lower than 10 km s$^{-1}$ (Brandi et al. \cite{brandi}), which is
consistent with the low velocity deduced for component \#3 ($\sim$6 km
s$^{-1}$). This establishes a further and independent link between this
feature and the circumbinary material. Brandi et al. associate
the cF-absorption system to the material streaming towards the hot
component. We emphasize that if component \#3 arises within this
gas, that conclusion does not hold, for this feature is still
visible on day +1.5, and it must therefore lie farther out. Indeed, at
this age the radius of the nova shell is $\approx$4.5$\times10^{13}
\mbox{cm}$, i.e. larger than the binary separation
($\approx$2$\times10^{13}$ cm).

%

From the spatially resolved HST observations of the 2006 eruption, it
is clear that the nova remnant has a bipolar morphology, resulting
from the interaction of the outburst ejecta with CSM that is
significantly denser in the equatorial regions of the binary than at
the poles (Bode et al. \cite{bode07}).  In general, the interaction of
the nova ejecta with the CSM should essentially modify the density and
velocity field of the CS environment formed during quiescence. The
accelerated CS gas presumably acquires a broad range of densities and
velocities, depending on the polar angle.  In this respect, it is
tempting to explain component \#1 as the outcome of the interaction
between the nova ejecta and the RG wind lost during the last quiescent
phase.

To provide a first order insight into this possibility, we consider a
spherical approximation, and assume that the mass $M_w$ is lost by the
RG with a wind velocity $u_w$ in the course of the last inter-outburst
phase (1986--2006).  The nova ejecta, characterized by mass $M_e$ and
kinetic energy $E_e$, sweep up the material lost during quiescence,
and forms a shell with mass $M_e+M_w$. This expands at a final
velocity $u$, which is dictated by momentum conservation. To estimate
this velocity one needs to express the initial momentum carried by the
ejecta as a function of $E_e$ and $M_e$. To this end, we assume that
the nova ejecta start to expand homologously, with an exponential
density distribution $\rho\propto \exp{(-v/v_0)}$, where the scale
velocity $v_0$ is determined by $E_e$ and $M_e$. Adopting typical
values $M_e=10^{-7}~M_{\odot}$ and $E_e=10^{43}$ erg, one gets
$v_0=9.1\times10^7$ cm s$^{-1}$ (note that the outer ejecta of course
have much larger velocities).  The ejecta momentum, integrated over the
sphere for the adopted ejecta structure, is $2M_ev_0$. Momentum
conservation then leads to the following relation between mass ratio
$\mu=M_e/M_w$ and velocities

\begin{displaymath}
\mu=\frac{u-u_w}{2v_0-u}\,.
\end{displaymath}

Adopting a typical RG wind velocity $u_w=20$ km s$^{-1}$, a final
velocity of the swept-up shell $u=50$ km s$^{-1}$, the ejecta
parameters $E_e=10^{43}$ erg and $M_e=10^{-7}~M_{\odot}$, from the
above expression one obtains $M_e/M_w=0.017$, or
$M_w\approx6\times10^{-6}~M_{\odot}$. The latter implies a RG
mass-loss rate $\dot{M}_{RG}$=$M_w/20\,\mbox{yr} \approx
3\times10^{-7}~M_{\odot}$ yr$^{-1}$, which is quite a sensible value
for a RG in a symbiotic system (Seaquist \& Taylor \cite{seaquist}).

Despite our crude approximations, this estimate indicates that
component \#1 might be related to the wind material accelerated by the
nova ejecta. If the absorption is produced by the shell formed by the
previous 1986 outburst, then the absorber with a velocity of $\sim$50
km s$^{-1}$ is located at a radius $\sim$3$\times10^{15}$ cm. In this
scenario, this spatial scale would also be equal to the distance
$\Delta r$ between subsequent shells. 

This simplified picture for component \#1 neglects two important
issues: i) the deviation of the CSM from spherical symmetry, which results
in the dependence of final velocities on the polar angle; ii) the
Rayleigh-Taylor instability of the swept-up shell, which results in
shell fragmentation.  The latter is essential for understanding the
absence of scattered emission in component \#1 (see above). Indeed, if
the covering factor of fragments is $C\ll1$, then the intensity of the
scattered emission is suppressed by the same factor $C$ with respect
to that arising in the spherical, smooth shell. On the other hand, the
absorption can be rather strong (i.e. comparable with that of the
spherical component) if the absorbing cloud accidentally gets into the
line of sight.

In fact, the absorption can be produced also by more distant CS clouds
formed by previous outbursts and mass-loss.  The \ion{Ca}{ii} column
density required to produce a \ion{Ca}{ii} $\lambda$3968 line with an
optical depth $\tau$=1 is $2.8\times10^{12}$ cm$^{-2}$ (for $b=10$ km
s$^{-1}$). If calcium is mostly \ion{Ca}{ii}, then for a solar
abundance this line can be detected, provided the cloud produced by
the shell fragmentation lies at a distance of about
2$\times10^{16}C^{-1/2}$ cm.  For instance, in the case of
$C\sim10^{-2}$, \ion{Ca}{ii} would be detectable up to a maximum
radius $r_{max}\sim2\times10^{17}$ cm.

Based on the available data, Iijima (\cite{iijima,iijima09})
identified the blue-shifted component at systemic velocity $-$33 km
s$^{-1}$ as possibly arising within the wind of the RG\footnote{Note
  that Iijima (\cite{iijima,iijima09}) used a systemic velocity of
  $-$35.8 km s$^{-1}$ for RS Oph. Adopting the value of Fekel et
  al. (\cite{fekel}), the systemic velocity of the blended \ion{Na}{i}
  component becomes $-$29 km s$^{-1}$.}. Our data, which have a
spectral resolution almost a factor 5 higher, show that multiple
components are present. As we have been arguing throughout this paper,
this is strongly suggestive of a more complex configuration, in which
the interaction between the recurrent outbursts and the RG wind plays a
fundamental role in shaping the CSM. Although we cannot make a firm
statement, it is possible that component \#2, characterized by a
systemic velocity of about $-$23 km s$^{-1}$, is the signature of the
RG wind. Another possibility is that it is the left-over of a previous
outburst, similarly to what we proposed for component \#1. At variance
with components \#1 and \#3, \#2 is still present in the \ion{K}{i}
profile after the outburst (see Fig.~\ref{fig:kievol}). This indicates
that the gas responsible for this absorption is placed at a
sufficiently large distance that its ionization balance is not too
much affected by the eruption.

\begin{figure}
\centerline{
\includegraphics[width=9cm]{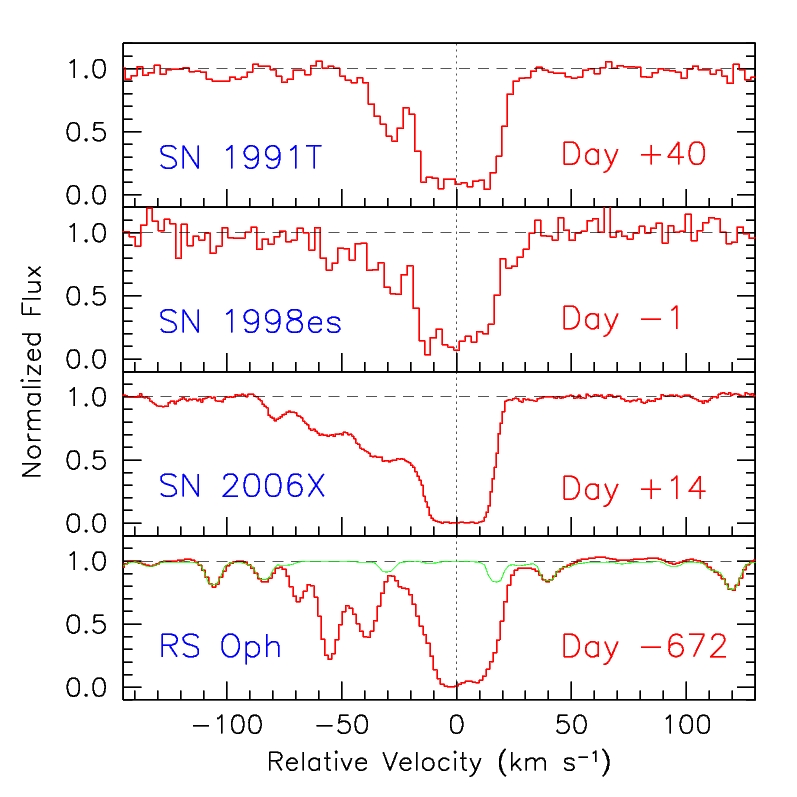}
}
\caption{\label{fig:blueshifted}Comparison between high-resolution
  spectroscopy of three Type Ia SNe and RS Oph in the region of
  \ion{Na}{i} D$_2$. For presentation the velocity scales refer to the
  center of the deep absorption, most likely associated to the local
  spiral arm. The data of SN~1991T, 1998es and 2006X are from
  Patat et al. (\cite{patat}).The light colored curve in the bottom 
  panel is an atmospheric model. }
\end{figure}

The presence of component \#3 after the nova outburst can be explained
by the resumption of the accretion, so that its re-appearance in our
latest epoch is not inconsistent with it being formed within the
circumbinary material ($r\leq$10$^{14}$ cm). On the other hand, the
presence of \#1 and \#2 at the same epoch is hard to explain if they
arise too close to the WD. Since the nova shell reaches a radius of
$\approx$10$^{15}$ cm around one month after the eruption, they would
have been swept away rather soon, had they been placed at shorter
distances. This suggests that the absorbing material is placed at a
sufficiently large radius that the ejecta are decelerated well before
reaching it.  In the light of these considerations, and given the
persistence of the \ion{Ca}{ii} H\&K and \ion{Na}{i} D lines during
the outburst, the most probable cause for the variability of
components \#1 and \#2 is photoionization.

If this is correct, then the recurrent outbursts do not destroy the
slow-moving shells produced by the previous events, implying that
systems like RS Oph are able to produce long-lasting structures in
their circumstellar environment. Structures that survive long enough
to show up during the most definitive and extreme event, when the WD
is finally incinerated by the thermonuclear runaway.

\section{\label{sec:concl}Conclusions}

Since the introduction of the mass-accreting WD scenario to explain
Type Ia SN explosions (Whelan \& Iben \cite{whelan}), the exact nature
of the progenitor system/s has been extensively debated (see Branch et
al. \cite{branch}; Livio \cite{livio}; Parthasarathy et
al. \cite{partha}; Tutukov \& Fedorova \cite{tutukov} for extensive
reviews). Among the various possibilities, symbiotic systems have been
proposed as a viable channel (Munari \& Renzini \cite{munari92}). In
particular, because of the mass of the WD hosted by RS Oph, which is
close to the Chandrasekhar limit, this system has been proposed as a
strong candidate (Hachisu \& Kato \cite{hachisu00},
\cite{hachisu01}). The detection of CSM material around SN~2006X and
2007le has strengthened this relation, showing the presence of
multiple, high-density, low-velocity clumps of neutral material along
the line of sight (Patat et al. \cite{patat}; Simon et
al. \cite{simon}).

The analysis presented in this paper, stimulated by the findings
reported by Iijima (\cite{iijima,iijima09}), leads us along the same
lines. If the presence of narrow, blue-shifted absorptions at
velocities where no interstellar component is expected is already an
indication that the material in which they originate has been lost by
the binary system, the time variability observed in coincidence with
the eruption definitely testifies in favor of its circumstellar
nature. In this scenario, the blue-shifted features would be explained
as arising in material swept by previous outbursts, as suggested for
SN~2006X (Patat et al. \cite{patat}) following the mechanism proposed
by Wood-Vasey \& Sokoloski (\cite{wood-vasey}), and hence buttressing
the link between the progenitor system of that particular SN and RS
Oph. In this respect, although based on a small number of objects, the
resemblance of RS Oph to a few Type Ia SNe is quite impressive
(Fig.~\ref{fig:blueshifted}). 

Of course, many issues remain open, like the exact mechanism that
produces the weakening of the absorption features, the location and
origin of the gas where the blue-shifted components arise
(inhomogeneities in the RG wind vs. relics of previous outbursts), and
so on. But, as in the case of SNe, the study of strong absorption
lines offers the unique possibility of investigating the presence of
low amounts of neutral gas, which would never be detected through the
typical signatures produced by direct interaction. Clearly, in the
case of a Type Ia SN, for which ejecta velocities are much larger, the
material generating component \#3 would be swept during the first
couple of days after the explosion, and hence it would be very hard
(if not impossible) to observe the same line variations we have
detected in RS Oph (see the discussion in Williams et
al. \cite{williams}, and Borkowski, Blondin \& Reynolds
\cite{borkowski} about the possible detection of narrow absorption
features by means of very early, high-resolution spectroscopy of Type
Ia. See also Hachisu, Kato \& Nomoto \cite{hachisu08} for
theoretical CSM scenarios in Type Ia progenitors).
 
This is why well time-sampled, high-resolution spectroscopy of
recurrent nova systems is crucial for better painting the complex
picture that is just emerging. The recent report by Sternberg et
  al. (\cite{sternberg}) of an excess of blue-shifted \ion{Na}{i}
  absorption features along the lines-of-sight to Type Ia supernovae
  might indeed find a physical background within this context.

\begin{acknowledgements}
This paper is based on observations made with ESO Telescopes at La
Silla Observatory under program IDs 073.D-0724(A), 076.D-0517(A) and
080.A-9207(A). F.P. acknowledges the kind hospitality of the
Department of Astrophysics (Oxford), and the Lorentz Center (Leiden),
where part of this work was conducted.

\end{acknowledgements}

\end{document}